\documentclass[preprint,12pt]{elsarticle}

\usepackage[nolist,nohyperlinks]{acronym} 
\usepackage{float}
\usepackage{graphicx} 
\usepackage{hyperref} 

\newcommand{\tab}{\hspace*{2em}}

\usepackage{algorithm}
\usepackage{algpseudocode}
\usepackage{algpascal}
\alglanguage{pseudocode}
\usepackage{color}
\algnewcommand{\LineComment}[1]{\State \(\triangleright\) #1}

\def\probP{\mathcal{P}}

\def\FC{\mathcal{F}}
\def\DC{\mathcal{D}}

\def\AC{\mathcal{A}}
\def\AWC{\AC}

\usepackage{amssymb}

\usepackage{color}
\journal{Information Fusion}

\begin{document}

\begin{frontmatter}



\title{\textit{SCARFF}: a Scalable Framework for Streaming Credit Card Fraud Detection with Spark \footnote{© 2017. This manuscript version is made available under the CC-BY-NC-ND 4.0 license http://creativecommons.org/licenses/by-nc-nd/4.0/}}


\author[1]{Fabrizio~Carcillo}
\author[1]{Andrea~Dal Pozzolo}
\author[1]{Yann-A\"el Le Borgne}
\author[2]{Olivier~Caelen}
\author[2]{Yannis~Mazzer}
\author[1]{Gianluca~Bontempi}

\address[1]{Machine Learning Group, Computer Science Department, Faculty of Sciences ULB, Universit\'e Libre de Bruxelles, Brussels, Belgium. \\(email: \{fcarcill, yleborgn, gbonte\}@ulb.ac.be, dalpozz@gmail.com)}
\address[2]{R\&D  High Processing \& Volume team, Worldline, Belgium. \\(email: \{yannis.mazzer, olivier.caelen\} @worldline.com).}

\begin{abstract}
The expansion of the electronic commerce, together with an increasing confidence of customers in electronic payments, makes of fraud detection a critical factor. Detecting frauds in (nearly) real time setting demands the design and the implementation of scalable learning techniques able to ingest and analyse massive amounts of streaming data. Recent advances in analytics and the availability of open source solutions for Big Data storage and processing open new perspectives to the fraud detection field. In this paper we present a SCAlable Real-time Fraud Finder (SCARFF) which integrates Big Data tools (Kafka, Spark and Cassandra) with a machine learning approach which deals with imbalance, nonstationarity and feedback latency. Experimental results on a massive dataset of real credit card transactions show that this framework is scalable, efficient and accurate over a big stream of transactions.
\end{abstract}


\end{frontmatter}

\begin{acronym}\addtolength{\itemsep}{-\baselineskip}
  \acro{AUC}{Area Under the receiver operating characteristic Curve}
  \acro{BRF}{Balanced Random Forest}
  \acro{BRT}{Balanced Random Tree}
  \acro{CP}{Card Precision}
  \acro{DDM}{Data Driven Model}
  \acro{EDR}{Expert Driven Rules}
  \acro{FDS}{Fraud-Detection System}
  \acro{GT}{Groups of Balanced Random Tree}
  \acro{NA}{Not Available}
  \acro{RDD}{Resilient Distributed Dataset}
  \acro{RF}{Random Forest}
  \acro{TBR}{Transaction Blocking Rule}
  \acro{SR}{Scoring Rule}
\end{acronym}
\section{Introduction}
\label{sec:intro}
The increasing adoption of electronic payments is opening new perspectives to fraudsters and asks for 
innovative countermeasures to their criminal activities. 
If on the one hand fraudsters continuously improve their techniques to emulate genuine behaviour, on the other hand
it becomes affordable for the companies managing transactional services to collect data about customers and monitor their behavior. 

The need of automatic systems able to detect frauds from historical data led to the design of a number of machine learning algorithms for fraud detection~\cite{ghosh1994credit,  sanchez2009association, sahin2013cost}. Supervised methods, typically based on binary classification, as well as unsupervised and one-class classification~\cite{Krawczyk201519,Krawczyk2015} have been proposed in literature. Most of these works address some specific issues of fraud detection, notably class imbalance~\cite{Krawczyk2016,dal2015When, dal201Calibrating} (the percentage of fraudulent transactions is usually very small), concept drift~\cite{Zliobaite2014,alippi2013JIT,Faria:2016:NDD:2883729.2883754,Abdallah2016,6597128,Yang2015158} (the distribution of fraudulent transactions might change in time) and stream processing \cite{41378,Lin:2015:SDS:2723372.2746485}.

The authors of this paper studied and analysed in detail the existing literature in previous works~\cite{dal2014learned, dal2014using, dal2015fraud, dal2016TNNLS} and proposed  an original solution for accurate classification of fraudulent credit card transactions in imbalanced and non-stationary settings. In particular we assessed the superiority of undersampling versus oversampling techniques in our specific problem, we proposed a sliding window approach to effectively tackle concept-drift and we addressed in~\cite{dal2015fraud, dal2016TNNLS} an issue often overlooked in literature: the \emph{verification latency} due to the fact that in real settings the transaction label is obtained only after that human investigators contacted the card holders. 

Though a large number of learning techniques have been proposed, most solutions assume a conventional setting where the entire dataset is resident in memory. It follows that very few studies made the implementation of these techniques scalable and studied their performances. Also what exists is typically related to other domains than the fraud: for instance~\cite{Ro2014OnTU} and~\cite{Triguero2015ROSEFWRFTW} studied already the issue of data imbalance in a Hadoop/MapReduce framework~\footnote{\url{https://hadoop.apache.org/docs/r1.2.1/mapred_tutorial.html}} but only for public and bioinformatics data. 

In domains closer to fraud detection most of the existing works are preliminary or in progress. H. Hormoz et al.~\cite{6620035} made a comparison between a serial implementation and a Hadoop/MapReduce batch processing  solution based on Artificial Immune Systems (AIS). The same authors made some tests on cloud services and provided accuracy measurements~\cite{6620034}. A web service framework for near real-time credit card fraud detection is described, together with some preliminary results, in~\cite{Tselykh:2015:WSD:2799979.2800039}. A big data architecture based on Flume, Hadoop and HDFS is proposed in~\cite{PatternAnalysis} but no validation results are provided.  An example of application in a non banking environment is presented in~\cite{Chen20151} where J. Chen et al. describe the Hadoop based fraud detection infrastructure at Alibaba. Other works in progress can be found on several git servers~\cite{github1,github2,github3,github4,github5,github6,github7}.

In this paper we start from the conclusions of our published works~\cite{dal2014learned, dal2015fraud, dal2016TNNLS}
and we propose a realistic and scalable implementation of a fraud detection system.
SCARFF (SCAlable Real-time Fraud Finder) is an open source platform which processes and analyses streaming data in order to return reliable alerts in a nearly real-time setting.
These are the main original contributions: 
\begin{enumerate}
    \item the design, implementation and test of an entirely open-source solution integrating state-of-the-art components from the Apache ecosystem. This architecture deals seamlessly with data ingestion, streaming, feature engineering, storage and classification;
    \item a scalable learning solution able to provide accurate classification in a context characterized by nonstationarity, class imbalance and verification latency. This is obtained by implementing in a scalable and distributed manner an ensemble solution able to deal with concept drift and delayed feedback;
    \item the design of a distributed on-line feature engineering functionality, which constantly updates historical features relevant to better identify fraud patterns. This on-line functionality relies on a MapReduce programming model;
\item a real-world extensive assessment, in terms of scalability, computational performance and precision, carried out by testing the platform on a stream of more than 8 millions of transactions (corresponding to more than 1.9 millions of cards) provided by our industrial partner;
\item the virtualisation of the complete workflow proposed in this article as a Docker container, making the workflow fully reproducible.
\end{enumerate}


The paper is organized as follows. Section \ref{sec:FDS} introduces the main characteristics of real-world Fraud-Detection Systems. Section \ref{sec:BD} gives an overview of the big data tools from the Apache ecosystem that are integrated in our framework. Section \ref{sec:SA} details the learning and the streaming functionalities of the platform.
Finally, in section \ref{sec:E} we assess the scalability, computational speed and precision on a real dataset, as a function of allocated resources and incoming transaction rates.


\section{Real-world Fraud Detection Systems}\label{sec:FDS}

\begin{figure*}[t!]
\begin{center}
    \includegraphics[width=1\textwidth]{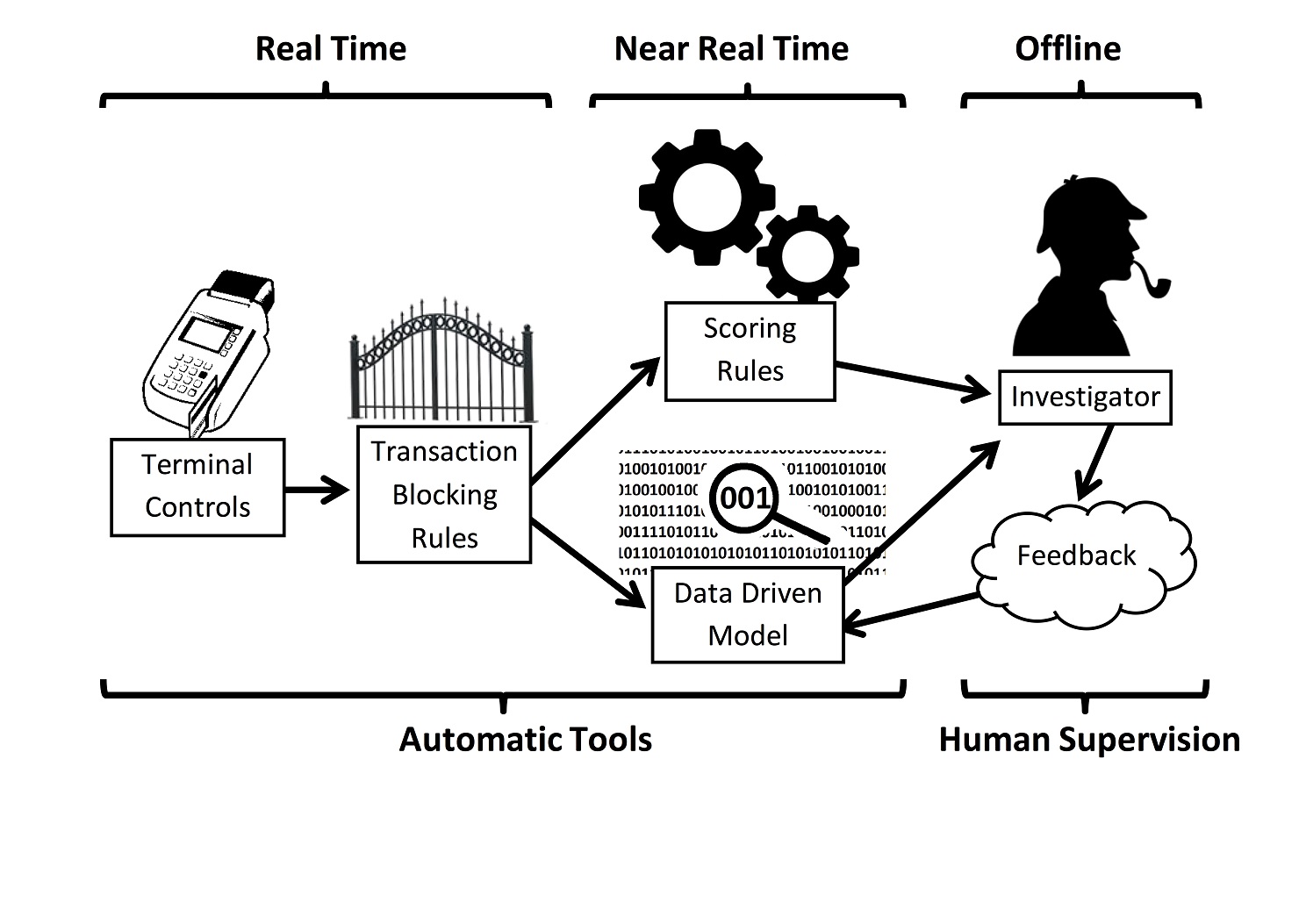}
    \caption{A diagram illustrating the layers of control in a \ac{FDS} \cite{veeramachaneni2016ai,dal2016TNNLS}. The paper focuses on the data driven model part.}
    \label{fig:FDS}
\end{center}
\end{figure*}

Real world \acp{FDS} for credit card transactions rely on both automatic and manual operations \cite{veeramachaneni2016ai,dal2016TNNLS} (Fig.~\ref{fig:FDS}).
Manual operations are performed offline by human investigators, while automatic components are implemented by algorithms that work in real-time and near real-time configurations.
Real-time operations take place before the payment is authorized, while near real-time operations are executed after the payment occurred.

Real-time processing consists of a set of security checks of the transaction. If these checks are not passed, the transaction is stopped, otherwise the amount is virtually transferred from one account to another.
Real-time operations can be divided in Terminal controls and \acp{TBR}.
Terminal controls concern terminal-card interaction (e.g. checking if the PIN code is correct) or terminal-server interaction (e.g. checking if there is a sufficient balance on the account). \acp{TBR} are a set of \emph{if/else} conditions, properly designed by fraud experts to block evident fraudulent attempts (e.g. \emph{IF attempt from a shop in black list THEN deny transaction}).
Those rules are seldom updated and because of their real-time nature (execution in milliseconds) they cannot rely on a feature engineering step returning complex features (e.g. cardholder profile or past cardholder behaviour).



Once the payment has been registered, near real-time operations are used to score transactions for fraud investigation.
Two near real-time kinds of control are typically performed: \acp{SR} and \ac{DDM}. 

\acp{SR} are expert based rules like the \acp{TBR}, but of a  more complex nature since they can take advantage of the output of a feature engineering step. For instance these rules can use the cardholders profile and behaviour (e.g. \emph{IF the cardholder is a 80 year old man who never used his card on-line AND the transaction involves a big amount AND the transaction comes from an offshore website THEN return a score of 0.9}).
The \acp{SR} output is a score obtained by merging the output of multiple rules and it is used to raise, if necessary, an alert on the transaction.
\acp{SR} are known to be effective for specific fraudster behaviours or recurrent fraudulent patterns.

The second near real-time component is the Data Driven Model (\ac{DDM}) which is based on Machine Learning classifiers trained to predict the probability of a new transaction to be fraudulent.  
The scalable implementation of this module, which is the main focus of this paper, will be detailed in section~\ref{sec:SA}.


The offline control layer is managed by investigators who take care of the \emph{alerts} returned by the \acp{TBR}, \acp{SR} and the \ac{DDM}.
By \emph{alert} we mean a transaction associated to high fraud risk and for which a human investigation is needed. 
In practice an alert is raised according to one of those criteria:
\begin{itemize}
\item the estimated risk of fraud associated to the transaction is over a threshold; 
\item the transaction belongs to the \emph{top-N} transactions
with the highest risk.
\end{itemize}
In the first case we may obtain an unpredictable number of alerts per day, while in the second case we may better organize the effort of the investigators by asking them to process alerts at a constant pace.

The number of credit cards a team of investigators can process depends on the organization guidelines, as well as the number of investigators available for such a task. Usually an organization investing a certain amount of money on investigation (i.e. employing a number of investigators), expects that at least a given number of alerts will be examined by its investigators. For this reason we have chosen to implement the second option in our pipeline. 

 

\section{The Big Data ecosystem}\label{sec:BD}

\begin{figure*}[!ht]
\begin{center}
	\includegraphics[width=1\textwidth]{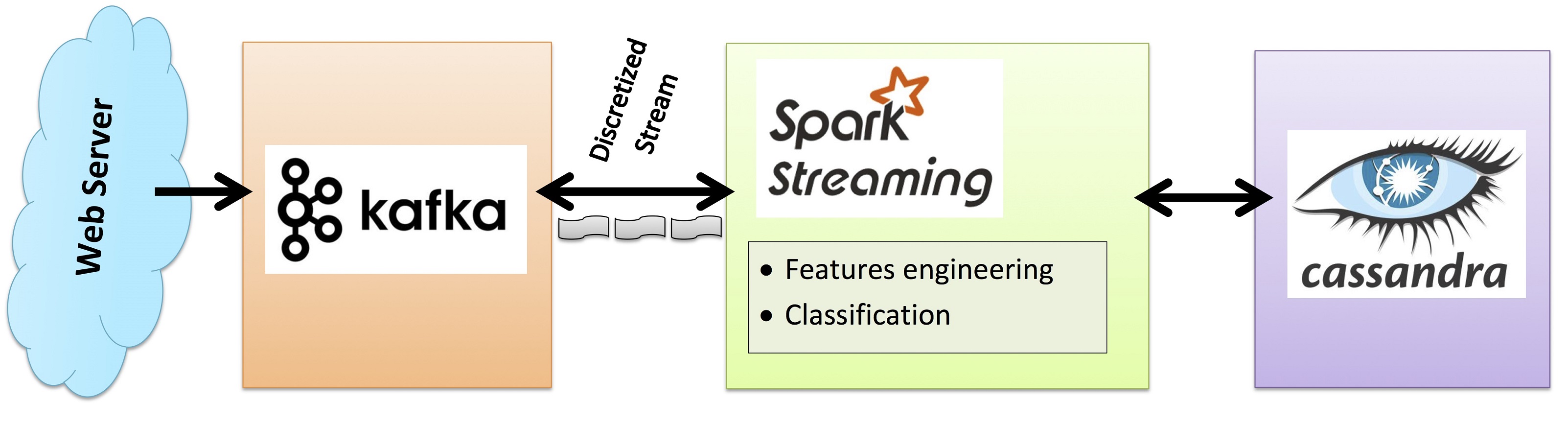}
    \caption{The Big Data pipeline. In our experimental setting, we used a bash program to emulate the web server which inputs data in the pipeline.}
    \label{fig:pipe}
\end{center}
\end{figure*}

This paper proposes a scalable implementation of the \ac{DDM} learning module
which relies on standard tools from the Apache ecosystem, notably Kafka, Spark and Cassandra (Fig.~\ref{fig:pipe}).
A major advantage of these components is that they similarly handle fault tolerance and tasks distribution.

\subsection{Transactions Collection}
\textbf{Apache Kafka}\footnote{\url{http://kafka.apache.org}} is a distributed publish-subscribe messaging queue system that is commonly used for log collection.
It has a multi-producers management system able to retrieve messages from multiple sources. 
For testing purposes we emulate the streaming through a bash program injecting transactions in Kafka at a desired rate per second. 
In case of need (e.g. system outage) the transactions may be retrieved also during a time interval (set by the user) posterior to their processing.
In general, data partitioning and retention make of Kafka a useful tool for fault tolerant transaction collection~\cite{kreps2011kafka}.

\subsection{Data Analysis}
\textbf{Apache Spark}\footnote{\url{http://spark.apache.org}} is an in-memory, streaming-enabled, Map-Reduce implementation which automatically distributes the computation among the assigned resources and aggregates the results on a distributed file system.
The central idea of this tool is to organize data in a distributed object, the \ac{RDD} \cite{Zaharia:2012:RDD:2228298.2228301}.
In case of partition lost, the RDD object contains sufficient information to retrieve the data structure \cite{Zaharia:2012:RDD:2228298.2228301,Zaharia:2010:SCC:1863103.1863113}.
Spark includes a built-in library for machine learning (package {\tt MLlib} \cite{meng2015mllib}), as well as one for streaming (package {\tt Streaming}).
A strong point of Spark is its capacity to enable batch and streaming analysis in the same platform.

The proposed framework relies on Spark Streaming which processes data stream in mini-batches trailing latency of the order of seconds. Though this could be considered as a disadvantage in some streaming contexts, it is harmless in our nearly real-time  setting.

The Spark module of the framework is written in Scala \cite{scala-overview-tech-report}, a language which combines object-oriented and functional programming. Scala runs on top of Java VM and it is fully compatible with Java libraries.
Overall, Spark accomplishes three missions in our pipeline: the aggregation of historical transactions to perform feature engineering, the online classification of the transactions returning the estimated fraud risk and the cold storage of transactions in Cassandra.

\subsection{Data Storage}
\textbf{Apache Cassandra}\footnote{\url{http://cassandra.apache.org}} is a distributed database designed for scalability, able to support replication across multiple nodes or datacenters.
It offers linear scalability, fault tolerance, low latency when querying \cite{Lakshman:2010:CDS:1773912.1773922}
and manages consistency of requests at the node level.
Data is stored on multiple nodes organized in a ring shape (i.e. there is no master and every node is as important as the others),
thus avoiding a single point of failure.
The creation of a Cassandra table requires the setting of some parameters (e.g. the primary key) having an impact on performances.
We use a compound primary key made of a partition and a clustering key.
The partition key is an identifier of the hour when the transaction has been received, making easy to retrieve old transactions and to compute statistics for a certain cardholder over a given period.
The clustering key defines the order of the records in a partition and it is composed of the card identifier and the timestamp.


\section{Online learning and streaming solutions}\label{sec:SA}

This section details the functionalities of the proposed framework.
Our pipeline implements two main functionalities: a machine learning classification engine and a streaming component.
In the first subsection, \ref{subsec:MLC}, the selected machine learning techniques are described.
The machine learning engine includes a weighted ensemble of two classifiers.
The second subsection, \ref{subsec:SA}, focuses on the streaming component.
Here, more details will be given regarding the data preprocessing (\ref{ssec:SA1}), the data throughput (\ref{ssec:DI}), the features engineering (\ref{ssec:SA3}), the online classification(\ref{ssec:SA4}) and the data storage (\ref{ssec:SA5}).




\subsection{The machine learning engine}\label{subsec:MLC}

This module is designed on the basis of our recent research in fraud detection~\cite{dal2015fraud, dal2016TNNLS} and it aims to take into account the specificity of a Fraud Detection System where automatic tools have to interact with human investigators.
The role of fraud investigators is to focus on the most suspicious transactions and to contact cardholders. This means that
the automatic system receives a binary feedback (fraud or genuine) only
on the small subset of transactions (few hundreds per day) which triggered an alert.  
For the rest of the transactions no feedback is received unless the cardholder reports a fraud. 
This means that non-alerted transactions can be assumed to be genuine only after some time.
The learning strategy discussed in~\cite{dal2015fraud, dal2016TNNLS} and implemented here, is able to integrate this verification latency by taking into consideration both transactions for which we have investigators' \emph{feedback} and those labeled by customer with some \emph{delay}.
In particular, the classification relies on Random Forests~\cite{breiman2001random, Rokach2016111}, which have been shown to be particularly effective in fraud detection problems ~\cite{bhattacharyya2011data, bahnsen2015example, van2015apate}.

The resulting algorithm estimating the risk of fraud  is then composed of two classifiers:
\begin{itemize}
\item a \textit{Feedback} Random Forest classifier $\FC_t$ trained on the observations generated in the last $f$ days and for which a \textit{Feedback} was returned by investigators;
\item a \textit{Delayed} classifier $\DC_t$ made of an ensemble of \acp{BRT} \cite{chen2004using,Rokach2016111} trained on the old transactions for which we can reasonably consider the class as known. Note that this classifier is typically learned on a much larger number of samples than the \textit{Feedback} one.
\end{itemize}

\begin{figure*}[!ht]
\begin{center}
	\includegraphics[width=1\textwidth]{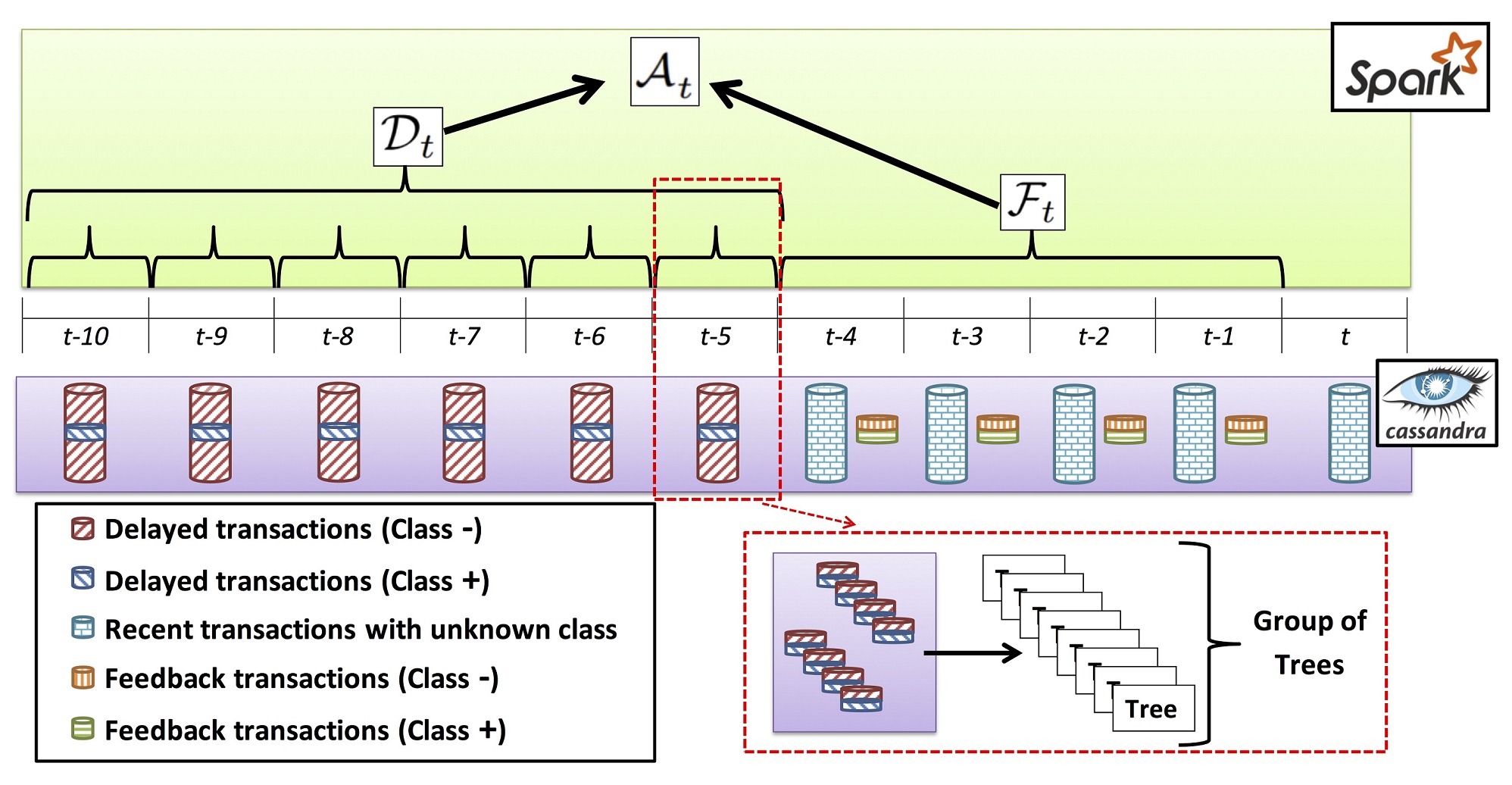}
    \caption{In this illustrative example, the Feedback model is 
    a Random Forest trained on investigator feedback from day $t-4$ to $t-1$. The Delayed model is an ensemble of Balanced Random Trees (\acp{BRT}), each trained on the observations of every single day from day $t-10$ to $t-5$. The transaction risk score is a function of the scores of the two models. Note that the \textit{Delayed} classifier follows a sliding window approach: as new \acp{BRT} are trained and added to the ensemble, the oldest ones are discarded. The same is true for the \textit{Feedback} classifier, where a sliding window approach is followed when selecting the subset of transactions used for the classifier training.}
    \label{fig:workingcond}
\end{center}
\end{figure*}

Every tree in $\DC_t$ is day specific, i.e. it uses only transactions of a given day (Fig.~\ref{fig:workingcond}). 
This allows an easier distribution of the computation and aggregation of the results.

Though the two classifiers are updated periodically (e.g. once per day), they are continuously used in the streaming module (subsection \ref{subsec:SA}) to assess the risk of fraud.

In order to deal with concept drift, a sliding window approach~\cite{dal2015fraud}
is used to update both $\FC_{t}$ and $\DC_{t}$ on the basis of new transactions.
The classifier $\FC_{t}$  is trained on the transactions (feedback) of the latest 14 days.
The classifier $\DC_{t}$ implements an updating strategy that keeps the \ac{BRT} corresponding to a window of 13 days and discards the oldest ones.

Given an incoming transaction $i$ at time $t$, coded by a feature vector $x_{i}$, the classifiers $\FC_t$ and $\DC_t$ produce respectively the posterior probabilities $\probP_{\FC_{t}}(+|x_i)$ and $\probP_{\DC_{t}}(+|x_i)$, where $+$ denotes a fraud and $-$ a genuine transaction.
The aggregated posterior probability $\probP_{\AWC_{t}}(+|x_i)$ is obtained by a weighted average of posterior probabilities from the individual classifiers: 
\begin{equation}
\label{eq:pat}
\probP_{\AWC_{t}}(+|x_i) = w^A\probP_{\FC_{t}}(+|x_i) + (1-w^A)\probP_{\DC_{t}}(+|x_i)
\end{equation}
where $w^A \in [0,1]$ and $\AWC_{t}$ is the overall model which wraps $\FC_{t}$ and $\DC_{t}$.
On the basis of the analysis conducted in~\cite{dal2015fraud} we set $w^A=0.5$.

The imbalanced nature of the classification problem led us to implement our own scalable version of a \ac{BRF}. For this purpose we integrated Scala code with Weka \cite{hall2009weka}, a well established open source tool for machine learning in Java.
The result is a scalable \ac{BRF} where every tree is trained on a subsample of the majority class (genuine cases) and the entire minority class (fraudulent cases). The pseudo-code of the scalable learner is detailed in \textbf{Algorithm \ref{alg:RandForest}}.

\newcommand{\nTrees}{nTrees}
\newcommand{\treesArrayGlob}{treeArrayGlobal}
\newcommand{\treesArray}{treeArray}
\newcommand{\frauds}{frauds}
\newcommand{\genuine}{genuine}
\newcommand{\subsetArray}{subsetGenuine}
\newcommand{\set}{balanced.set}
\newcommand{\tree}{balanced.tree}
\begin{algorithm}
\caption{Distributed implementation of a Balanced Random Forest}
\label{alg:RandForest}
\begin{algorithmic}[1] 
\State $\nTrees\gets$ number of trees per partition
\State $\frauds\gets$ array of frauds
\State Broadcast $\frauds$
\State $\genuine\gets$ RDD of genuine
\algrenewcommand\algorithmicwhile{\textbf{for any partition}}
\While{($\genuine$)} \label{row:forp}
\State $\treesArray\gets$ initialize an array
\For{$i\gets 0,\, \nTrees-1$}\label{row:fornt}
\State $\subsetArray \gets$ random subsample of the partition
\State $\set\gets Union(\frauds, \subsetArray)$ 
\State $\tree\gets $ build a classifier using $\set$ \label{row:efornt}
\State $\treesArray\gets $ append $\tree$ to $\treesArray$ 
\EndFor 
\EndWhile
\State $\treesArrayGlob\gets$ collect all $\treesArray$ from partitions
\end{algorithmic}
\end{algorithm}

Note that  the genuine transactions are stored in the \ac{RDD} $\genuine$  while all the fraudulent transactions (array  $\frauds$) are broadcast\footnote{In Spark, the broadcast mechanism allows to keep a read-only variable cached on each machine rather than shipping a copy of it with tasks. They can be used, for example, to give every node a copy of a large input dataset in an efficient manner.} to every executor.
For any partition of the RDD $\genuine$  we build $\nTrees$ \acp{BRT}
and we  collect them in $\treesArray$. Finally we store all the models created in the partitions
in the $\treesArrayGlob$ object.
\textit{Delayed} classification task is then performed by aggregating the outcome of all the \acp{BRF}.
A typical way to perform aggregation  relies on weighting 
\begin{equation}
\label{eq:weight}
\probP_{\DC_{t}}(+|x_i) = \sum_{n=1}^{k} w^D_n \probP_{BRF_{t-d-n}}(+|x_i)
\end{equation}
where $w^D$ is a vector of $k$ weights which sums to 1 and $d$ is the delay (number of days) for the reception of the labels. Different strategies can be used to set the weights, e.g. proportionally to the size of the tree training set or to the number of incorrect decisions (Dynamic Weighted Majority). 
A better ensemble learning strategy may be used to optimize the detection task.
To further investigate ensemble strategies for streaming classification, we suggest the read of these surveys \cite{Wozniak20143,krawczyk2017ensemble,gomes2017survey}.

\subsection{The streaming analytics engine} \label{subsec:SA}
This engine implements the following functionalities:
\begin{itemize}
\item data throughput;
\item data preprocessing;
\item features engineering;
\item online classification;
\item data storage.
\end{itemize}

\subsubsection{Data throughput} \label{ssec:DI}
\begin{figure*}[t!]
\begin{center}
	\includegraphics[width=1\textwidth]{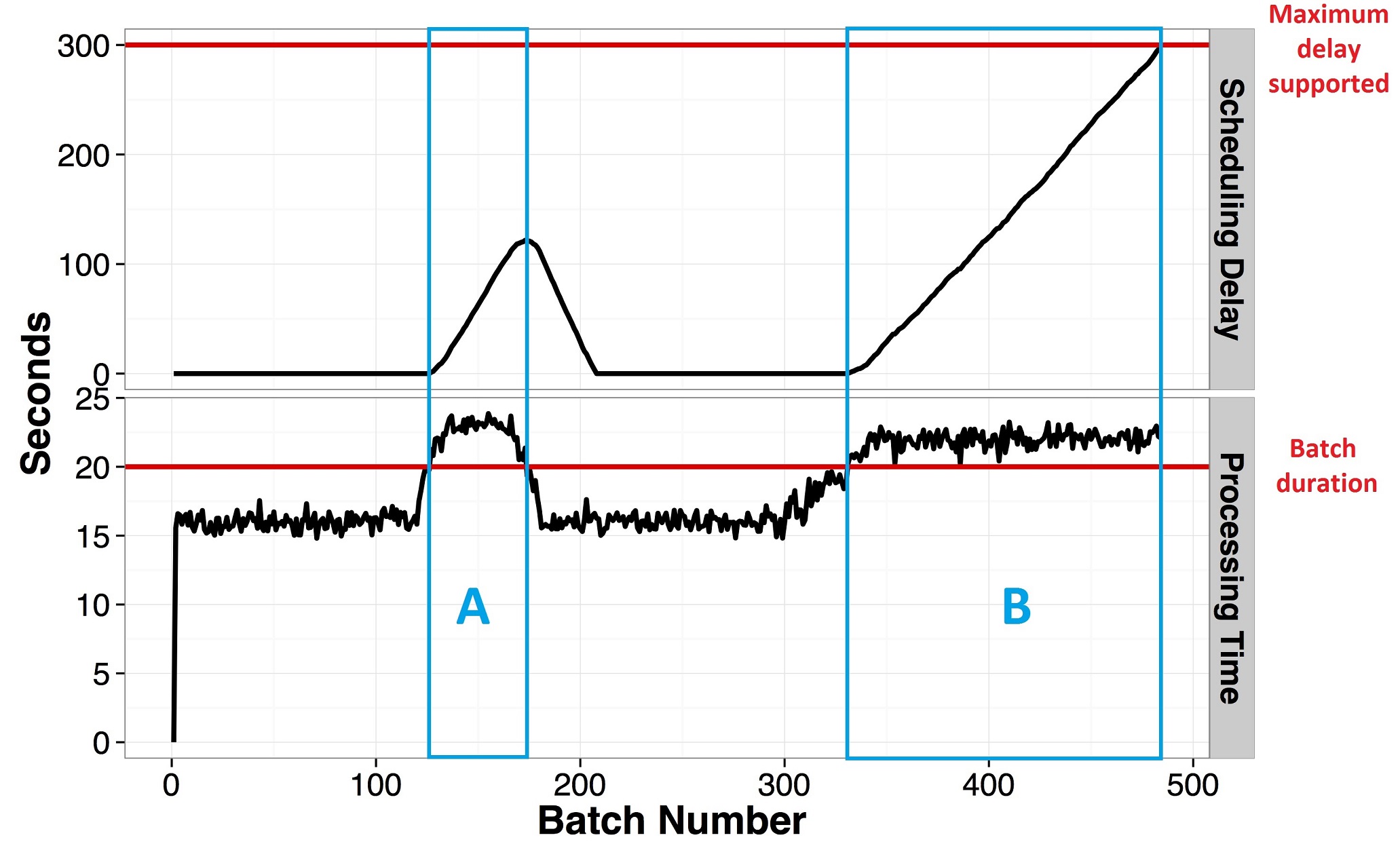}
    \caption{Behavior of the scheduling delay vs. batch duration in a synthetic example where the available resources can afford a maximum delay of 300 seconds. Configuration A refers to a situation where, only  for a limited period, processing time is longer than the batch duration. Configuration B
    corresponds to a longer violation period of maximal batch duration. This event leads to an increase of delay up to a limit imposed by the available resources.}
    \label{fig:delay}
\end{center}
\end{figure*}

Data throughput in Spark from Kafka produces a DStream object (Discretized Stream) \cite{Zaharia:2012:DSE:2342763.2342773}, the basic abstraction provided by Spark Streaming to represent a continuous stream of data. A DStream object is a continuous series of \acp{RDD}, obtained by  periodically generating and appending \acp{RDD}. 
The frequency at which streaming data are partitioned into batches (a.k.a. batch duration) is an important parameter of a DStream object.
In fact, the processing of a new batch starts as soon as a new \ac{RDD} is generated and the processing of the previous batch  has been completed.
This entails that the processing time of an \ac{RDD} should be smaller than the batch duration. 
If this is not the case, the batch is stacked in a queue and the execution postponed. Such a delay is sustainable for a limited period of time only (see configuration A in Fig.~\ref{fig:delay}). 
If incoming data flow at a too high rate for a long period the application fails as soon as all the storage resources are exhausted (see configuration B in Fig.~\ref{fig:delay}).

\subsubsection{Data Preprocessing}\label{ssec:SA1}
This step deals with the treatment of missing values  (replaced by median values) and with the coding of the categorical features (e.g. product class or merchant business type) characterized by a large number of values. 
The coding step consists in replacing each categorical value by a numeric value representing the a priori probability of the category to be associated to a fraudulent transaction, as presented in~\cite{friedman2001elements} (subsection 9.2.4). The probability is estimated from
historical data and stored  in a \textit{dictionary}~\cite{dal2015adaptive}.

The effectiveness of such preprocessing is confirmed by previous research as well as by our industrial partner experience. From a more theoretical perspective it can be seen as an instance of \emph{cascade generalization} \cite{Gama2000} where preliminary naive classifiers are used as inputs to a more powerful classifiers. Potential risks of concept drift in this procedure could be addressed by updating the dictionary every time a new batch of labels is received.
A detailed survey on data preprocessing for data stream mining can be found in~\cite{ramrez2017survey}.

\subsubsection{Feature engineering}\label{ssec:SA3}

This step consists in the retrieval of historical data stored in the Cassandra database and the computation of aggregated statistics. Commonly used statistics  are the maximum, minimum, count and average of relevant numerical variables (e.g. transaction amount), which derive from recent transactions of the concerned cardholder.
In this step, crucial parameters are the size of the historical time window (e.g. week or month) and the number of recent transactions taken into consideration. 
Given the streaming nature of the problem, the modification of these parameters can noticeably impact the required resources and the affordable rate of data throughput.
Alternative feature engineering techniques are discussed in literature~\cite{CorreaBahnsen2016134, Whitrow:2009:TAS:1485071.1485090}. 

\subsubsection{Online classification}\label{ssec:SA4}
This step consists first in classifying any incoming transaction by using the most recent classification model returned by the procedure described in subsection~\ref{subsec:MLC}.
Once the classification is performed, the system updates a dashboard containing a priority list of transactions (alerts) sorted by estimated risk. In our prototype this dashboard is simply a database table. Obviously a more user-friendly interface should be considered in a production environment. 

\subsubsection{Data storage}\label{ssec:SA5}
The final step consists in storing transactions and their aggregated information in a Cassandra table by means of a Spark Cassandra Connector, an open source library developed by Datastax \footnote{\url{https://github.com/datastax/spark-cassandra-connector}}.
Transaction and aggregated features are periodically retrieved to build the training set of the machine learning engine.

\newcommand{\dStream}{DStream}
\newcommand{\inTrx}{inTrx}
\newcommand{\inTime}{currentDate}
\newcommand{\tableTrx}{tableTrx}
\newcommand{\tableRank}{tableRank}
\newcommand{\arrRank}{AlertTable}
\newcommand{\topalerts}{topN}
\newcommand{\histTrx}{featHist}
\newcommand{\idTrx}{idTrx}
\newcommand{\window}{window}
\newcommand{\modTime}{modelDate}
\newcommand{\augTrx}{augTrx}
\newcommand{\feedProb}{feedProb}
\newcommand{\delProb}{delProb}
\newcommand{\totalProb}{totalProb}
\newcommand{\trainFeed}{trainFeedback}
\newcommand{\trainDel}{trainDelayed}
\begin{algorithm}
\caption{Streaming procedure}
\label{alg:SparkStreaming}
\begin{algorithmic}[1] 
\State $\dStream\gets$ RDD collection
\State $\tableTrx\gets$ empty cassandra table
\State $\tableRank\gets$ empty cassandra table
\State $\arrRank\gets$ empty array
\State $\topalerts\gets$ number of alerts to retain
\State $\window\gets$ array of window intervals for\\
\tab \tab \tab features aggregation
\State $\modTime\gets$ day of the last model update
\algrenewcommand\algorithmicwhile{\textbf{for any RDD}}
\While{($\dStream$)}
\State $\inTrx\gets$ current RDD 
\State $\inTime\gets$ date of $\inTrx$
\If {$\inTime \neq \modTime$} \label{row:iftime}
\State $\trainFeed\gets$ train a new $\FC$
\State $\trainDel\gets$ train a new $\DC$
\State $\modTime\gets \inTime$ 
\State $\tableRank\gets \arrRank$ 
\State $\arrRank\gets$ empty array
\EndIf 
\State $\idTrx\gets getUniqueIds(\inTrx)$
\State $\histTrx\gets$ empty array
\For{$i\gets 0,\, size(\window)-1$} \label{row:forinterval}
\State $h\gets retrieveHist(\idTrx,\tableTrx,\window(i))$
\State $\histTrx\gets$ append $h$ to $\histTrx$
\EndFor
\State $\augTrx\gets$ merge $\inTrx$ and $\histTrx$
\State $\tableTrx\gets$ insert $\augTrx$
\State $\feedProb\gets$ classify $\augTrx$ using $\trainFeed$ and get the probability for class fraud \label{row:beg}
\State $\delProb\gets$ classify $\augTrx$ using $\trainDel$ and get the probability for class fraud
\State $\totalProb\gets$ ensemble $\delProb$ and $\feedProb$
\State $\arrRank\gets$ append $\totalProb$ to $\arrRank$ and keep the $\topalerts$ alerts with the highest risk\label{row:end}
\EndWhile
\end{algorithmic}
\end{algorithm}

\subsubsection{Pseudo-code}
The entire streaming procedure can then be summarized by the pseudo-code in \textbf{Algorithm \ref{alg:SparkStreaming}}.
Given a $\dStream$, for any of its \ac{RDD} components (denoted $\inTrx$) we perform a series of tasks:
\begin{itemize}
\item if the day is over [Row:\ref{row:iftime}], we retrain the models $\FC$ and $\DC$ (section~\ref{subsec:MLC}), we save on Cassandra DB the $\topalerts$ alerts from $\arrRank$, we reset $\arrRank$ and we discard the unneeded transactions;
\item for any given time interval in the array $\window$, we retrieve information about previous transactions from $\tableTrx$ and for any cardholder ($\idTrx$) [Row:\ref{row:forinterval}];
\item once the feature vector $\histTrx$ is built, the transaction may be classified according to the up-to-date classifiers $\FC$ and $\DC$ [Rows:\ref{row:beg}-\ref{row:end}] and the riskiest $\topalerts$ alerts stored in the alert table.
\end{itemize}


\section{Experiments}\label{sec:E}

This section assesses the proposed scalable architecture according to different criteria: 
\begin{itemize}
  \item Scalability;
  \item Impact of internal parametrization on computational performance;
  \item Classification precision.
\end{itemize}
Experiments were carried out on a cluster of ten machines, each with 24 cores and 80GB of RAM.
Spark was run on top of the cluster resource manager Yarn \cite{Vavilapalli:2013:AHY:2523616.2523633}.
For all experiments, each executor was allocated 1GB of RAM, and the driver was allocated 10GB of RAM. Further discussion over memory usage will be presented in this section.

The dataset $DS$ used for experiments is a selection of 40 consecutive days of transactions recorded from 2014, October, 18 to November, 26. This dataset includes more than 8 millions of e-commerce transactions from almost 2 millions cardholders, 18 descriptive features and the label (genuine or fraudulent).
Table \ref{table:data} reports the presence of frauds in terms of fraudulent transactions and fraudulent cards.
Note that the feature engineering step is performed on a one week time window leading to the creation of 17 additional features.

The \textit{Feedback} classifier $\FC_{t}$ is trained over all the transactions from the 100 cards alerted per day and for a period of 14 days. Given that there are on average four transactions per card per day, $\FC_{t}$ is trained with about 5,600 transactions.

The \textit{Delayed} classifier $\DC_{t}$ is trained on the set of transactions (about 2.7 million) occurring during 13 days (from day $t-8$ to $t-20$). The total number of transactions can be roughly estimated as follows 
$$2.4 \frac{trx}{sec} \times 86,400 \frac{sec}{day} \times 13 days=2,695,680trx$$
Note however that the effective size of the final training set is smaller and dictated by the undersampling step
which returns a more balanced dataset.

Note also that in the experiments we use a weighting strategy for aggregation~(Equation \ref{eq:weight}) where weights are proportional to the number
of training samples per tree. 

\begin{table}[!t]
\centering
\begin{tabular}{|l||r|r|r|r|}
 \hline
 \multicolumn{5}{|c|}{Data subset} \\
 \hline
 Subset name & \# trx & \% of fraud. trx & \# cards & \% of fraud. cards\\
 \hline

 $DS$ & 8,356,811  & 0.4 & 1,921,457 & 0.2\\
 \hline
\end{tabular}
\caption{Dataset used for experiments}
\label{table:data}
\end{table}

\subsection{Scalability}
This section aims to assess the scalability of our pipeline by running three times the entire detection procedure on the dataset $DS$ with an increasing number of Spark executors (25, 35 and 45).
We set the batch duration \footnote{This is an attribute of the object {\tt StreamingContext} in Spark} to 240 seconds (section \ref{ssec:DI}) and the data incoming rate at 100 transactions per second (trx/sec). Note that the real throughput rate associated to $DS$ is 2.4 trx/sec.

\begin{figure*}[!t]
\begin{center}
    \includegraphics[width=1\textwidth]{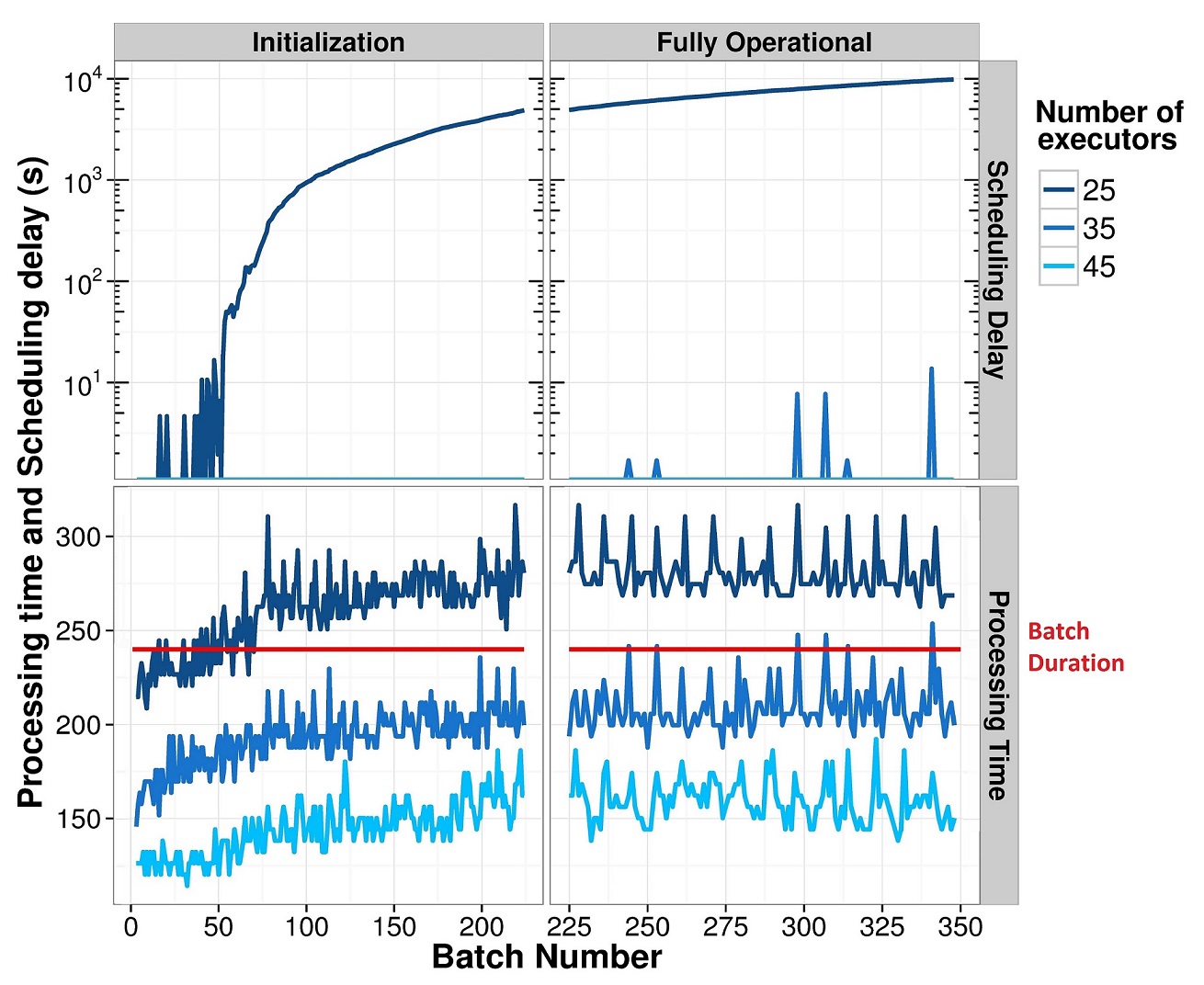}
    \caption{Processing time (lower part) and scheduling delay (upper part) behavior as a function of the number of executors. The left facet focuses on the \emph{Initialization} phase, while the right facet reports the behaviour during the \emph{Fully Operational} phase.  On the x-axes we have an in time ordered sequence of batches analysed by the application, while on the y-axis we have the number of seconds needed to process a given batch.}
    \label{fig:lines}
\end{center}
\end{figure*}

Fig.~\ref{fig:lines} is divided in two vertical facets, displaying the execution behavior during an \emph{Initialization} phase and a \emph{Fully Operational} phase, respectively. During the \emph{Initialization} phase, the system is in a bootstrap state, the Cassandra database is not completely filled and classification is not yet started. By \emph{Fully Operational} phase we mean that all the functionalities (preprocessing, feature engineering and classification) are fully working. 
Fig.~\ref{fig:lines} shows that during the \emph{Initialization} phase, the processing time increases because of (i) the growing number of stored transactions and (ii) the increasing classification time due to the growing complexity of the random forest used for classification.

The \emph{Fully Operational} phase begins as soon as the number of days set for features engineering is elapsed and we reach the desired number of models in the ensemble. From this moment on, the learning system starts discarding the oldest transactions and the oldest models from the ensemble, thus keeping constant the memory occupation.

Considering the \emph{Fully Operational} phase, a first observation is that the processing time for the \emph{25 executors} run is longer than the limit set by the batch duration. 
In this case a delay will be accumulating with a potential risk of application failure (see also Fig.~\ref{fig:delay}).

This is not the case for 35 and 45 executors, respectively, since the processing time is typically shorter than the batch duration. 
Nevertheless in the \emph{35 executors} case, we still observe some peaks passing over the batch duration threshold.
Those peaks refer to batches where the retraining of the model takes place in addition to the feature engineering and the online classification operations. 
The fact that some distributions pass over the 240 seconds threshold has not necessarily a negative impact on the resulting performances. This is due to the fact that in real production setting, these situations occur only once a day with minor impact on final schedules.

\begin{figure*}[!ht]
\begin{center}
    \includegraphics[width=1\textwidth]{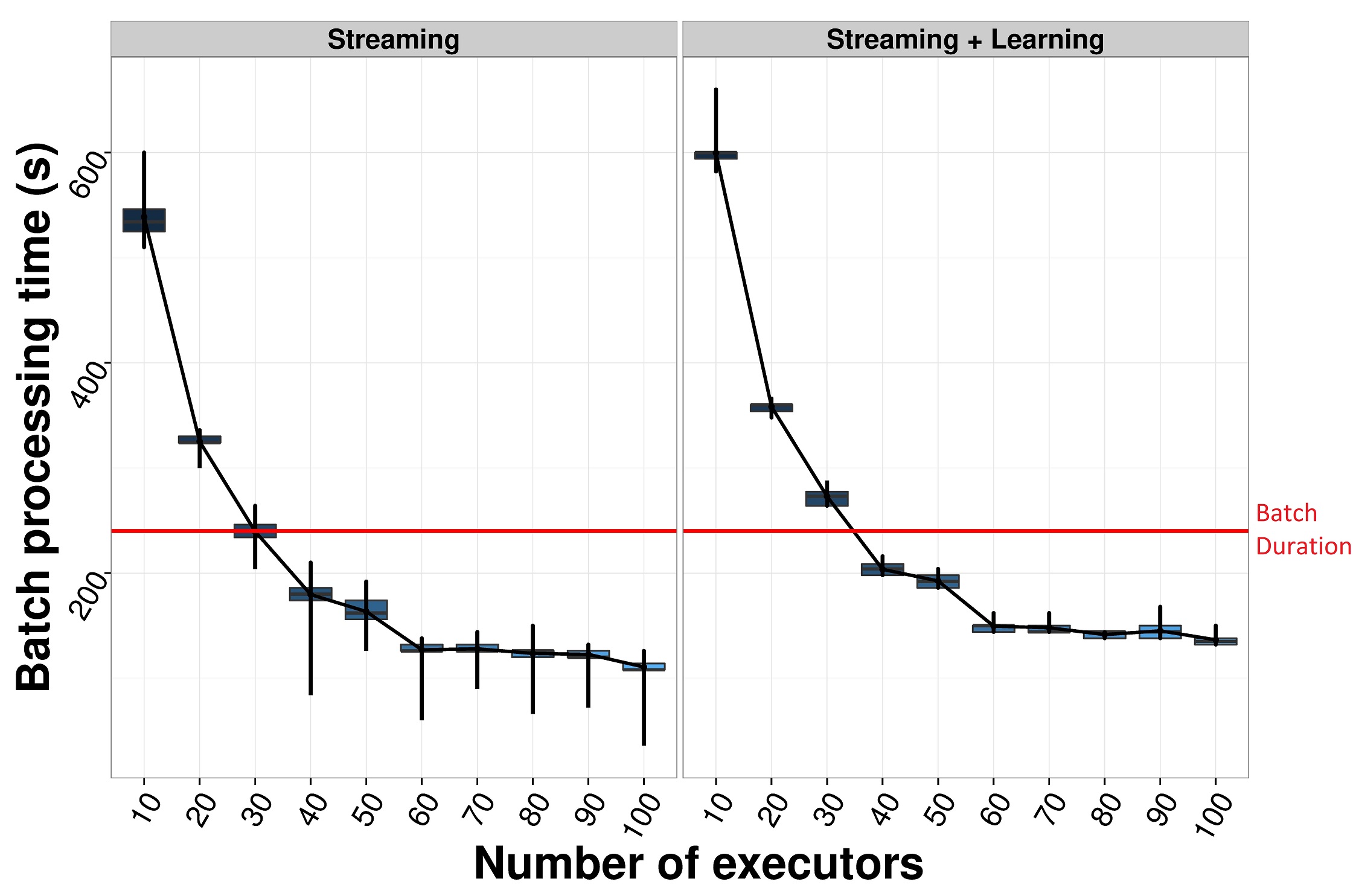}
    \caption{Processing time distribution in the \emph{Fully Operational} phase, for different number of executors. Left: execution time due to \emph{Streaming} only. Right: execution time due to \emph{Streaming} and \emph{Learning}.}
    \label{fig:boxplot}
\end{center}
\end{figure*}

For this reason we have rearranged the \emph{Fully Operational} data in Fig.~\ref{fig:boxplot} in order to make explicit the processing time due to the streaming and to the learning step, respectively. It appears again evident that the configuration with 10, 20 or 30 executors is not sufficient to absorb a streaming rate of 100 trx/sec.

Let us remark also that in Fig.~\ref{fig:boxplot} the average processing time is decreasing with the number of executors suggesting that the application is scalable. However the decreasing rate of improvement suggests that the improvement could be negligible from a certain number of executors on. This is typically due to the fact that the map-reduce process may become too expensive for a large number of executors since the benefit of dividing the computation among executors is counterbalanced by the cost of shuffling too many data among those executors.

\subsection{Impact of internal parametrization on computational performance}
In the previous section, we remarked that it is possible to define the minimal number of executors able to manage a given incoming  rate (e.g. 100 trx/sec). However, given the potential saturation of a distributed approach, it is interesting to study how to deal with high incoming rates without necessarily increase the size of the cluster. A possible solution comes from an appropriate tuning of internal parameters like the batch duration time. However an increase of the batch duration time implies two drawbacks:
\begin{itemize}
  \item a deterioration of the precision of the features engineering step;
  \item a delay in raising alerts.
\end{itemize}
The precision of features engineering is reduced since during its calculation, only the transactions stored in advance may be used; therefore if we have two transactions from the same card in a given batch, information about the first transaction will not be included to engineer features related to the second one.

The second drawback is a minor one since, as discussed previously, in a fraud detection scenario where human investigators have to contact clients to obtain their feedback, a delay of few minutes makes little difference.

\begin{figure*}[!ht]
\begin{center}
    \includegraphics[width=1\textwidth]{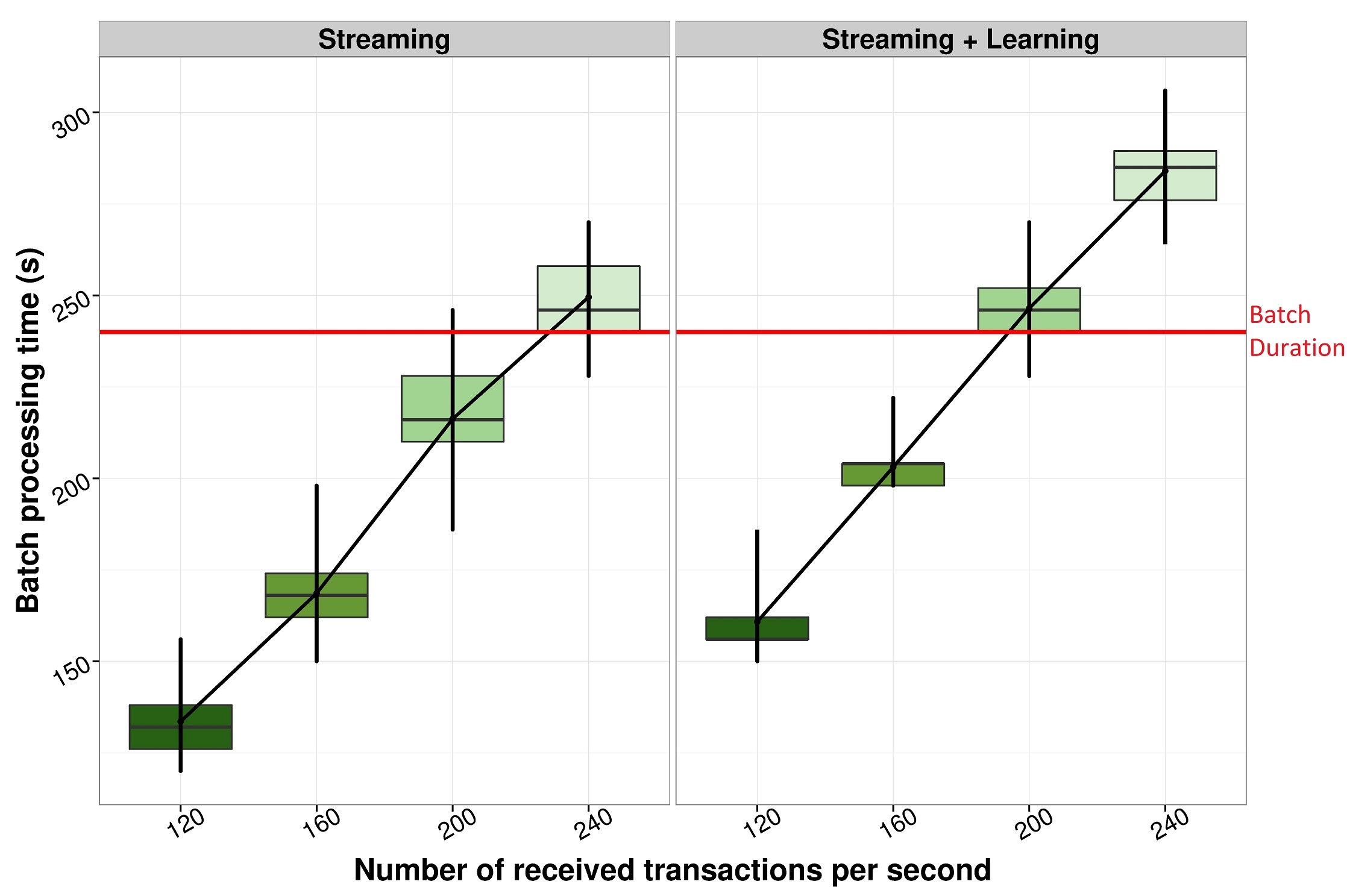}
    \caption{Processing time distribution in the \emph{Fully Operational} phase for different throughput rates, fixed batch duration of 240 seconds and using 100 executors. The left side shows the distribution of \emph{Streaming} times while the \emph{Streaming + Learning} times are shown on the right side. }
    \label{fig:speed}
\end{center}
\end{figure*}

In Fig.~\ref{fig:speed} we report the results obtained with the dataset $DS$, 100 executors and by raising the incoming rate to 240 trx/sec. The aim is to test the robustness of the infrastructure for long periods at high throughput rates.


It is worth to notice that 200 trx/sec is not necessarily an upper limit and that still higher rates could be obtained either by increasing the batch duration or the number of executors.

In terms of RAM-Hours (the average RAM usage in gigabyte per hour required by a given process~\cite{bifet2010fast}), our solution exhibits two distinct behaviours. In the \emph{Initialization} phase, the memory use grows linearly in time until the \emph{Fully Operational} phase. In the \emph{Fully Operational} phase, memory use is constant (0.05 RAM-Hour in case of a 200 trx/sec stream and 100 executors).

\begin{figure*}[!ht]
\begin{center}
    \includegraphics[width=1\textwidth]{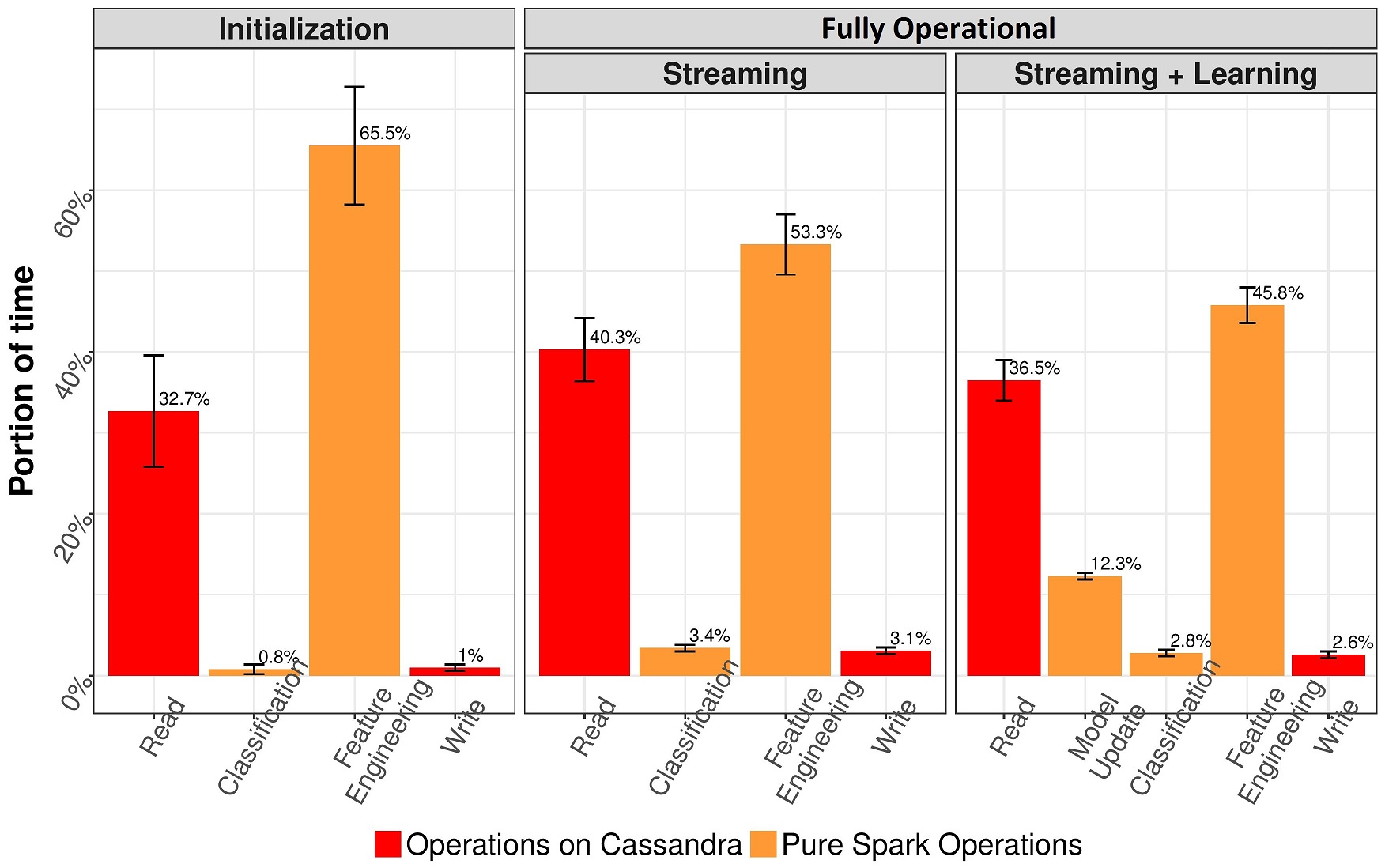}
    \caption{Distribution of processing time among \emph{Reads} and \emph{Writes} on Cassandra, \emph{Feature engineering}, \emph{Model Update}  and \emph{Classification} tasks.}
    \label{fig:timeusage}
\end{center}
\end{figure*}

Another interesting information concerns how the processing time is distributed among the different tasks of the Streaming functionality.
From Fig.~\ref{fig:timeusage}, it appears that the heaviest task is the \emph{Feature Engineering} (which includes the aggregation described in \ref{subsec:SA}), followed by the reading time from Cassandra.

The distribution of \emph{Read} operations and \emph{Write} operations is strongly skewed in our plot.
That is happening because we are writing few lines and reading many lines.
Indeed for the aggregation purposes, we need to retrieve old transactions information and this concerns far more lines than the one pushed in Cassandra.
As expected, the \emph{Read} tasks as well the \emph{Classification} tasks, consume more resources in the \emph{Fully Operational} phase than in the \emph{Initialization} one. Note that the time for \emph{Feature Engineering} is similar during the two phases in absolute values: the decrease visible in the figure is only in percentage terms (\emph{Read} and \emph{Write} times increase).
The \emph{Fully Operational} phase includes two sub-phases (\emph{Streaming} and \emph{Streaming + Learning}).
An additional component, the \emph{Model Update}, characterizes the latter sub-phase and impact for the 12.3\% of the total processing time.

\subsection{Classification Precision}

Fraud Detection Systems are designed to have accurate detection performance. A good measure for the precision, proposed in \cite{dal2016TNNLS} and previously used in rare item detection \cite{fan2011detection}, is the \ac{CP}, which is the proportion of detected fraudulent cards among the alerted ones.

\begin{figure*}[!ht]
\begin{center}
    \includegraphics[width=1\textwidth]{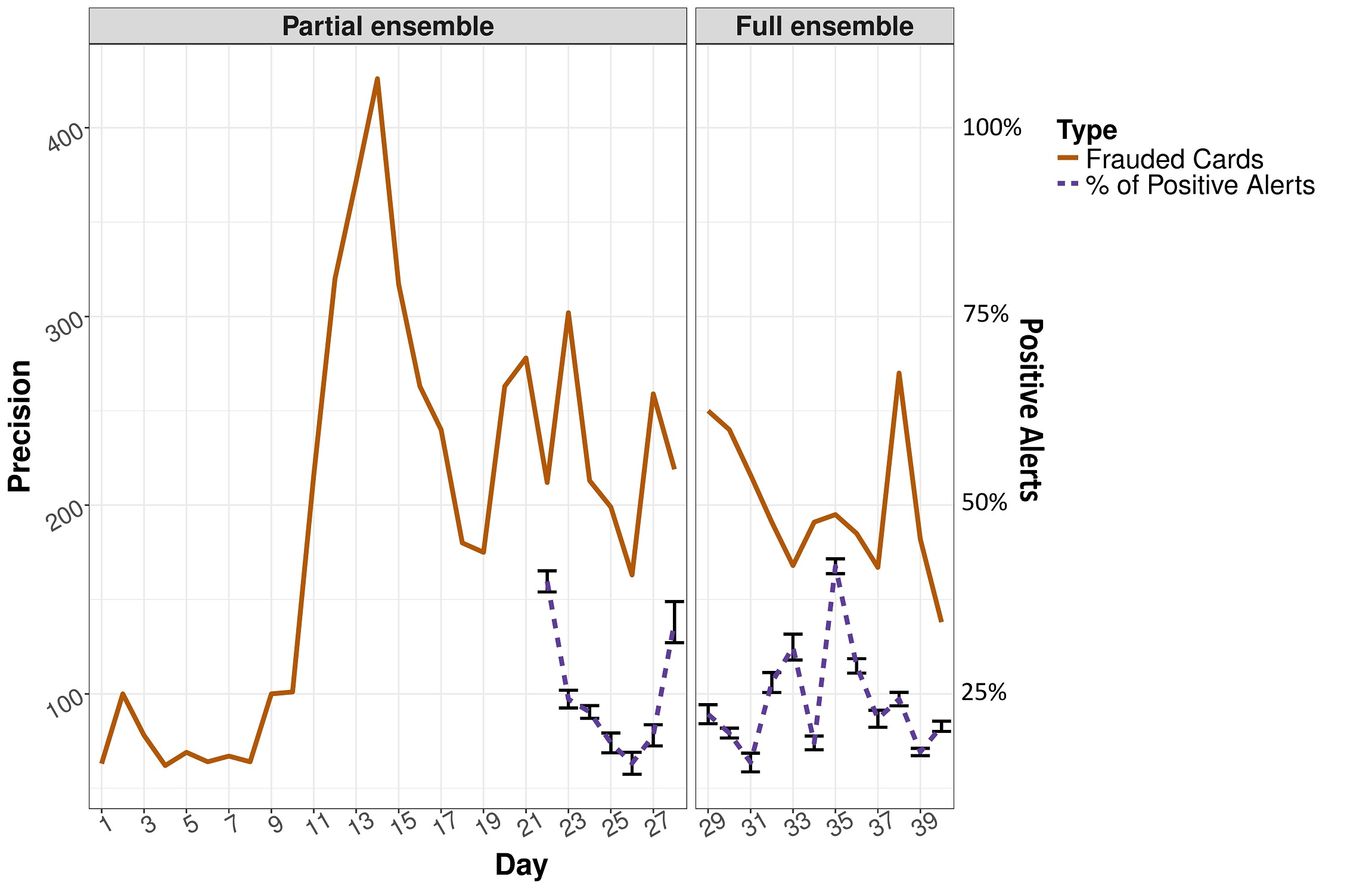}
    \caption{The line chart shows the distribution of frauded cards (solid/yellow line, left y-axis) and the percentage of detected frauds in the examined period (dashed/violet line, right y-axis). The confidence intervals were obtained by repeating the experiment ten times and changing different random seeds.}
    \label{fig:fraudsperday}
\end{center}
\end{figure*}

It is therefore important to assess the \ac{CP} returned by a scalable implementation of the fraud detection procedure.
The precision obtained in our experiments with the dataset $DS$ is essentially in line with the results published in \cite{dal2016TNNLS}. The minor discordance is due to the fact that we are using only a subset of the features used in \cite{dal2016TNNLS}.

Overall we obtained an average precision $CP_k = 0.24$ where $k$ is set to 100 since this is the average number of cards that can be daily checked  by the investigators working for our industrial partner. This means that on average 24 alerts out of 100 are correct.

\begin{figure*}[!ht]
\begin{center}
    \includegraphics[width=1\textwidth]{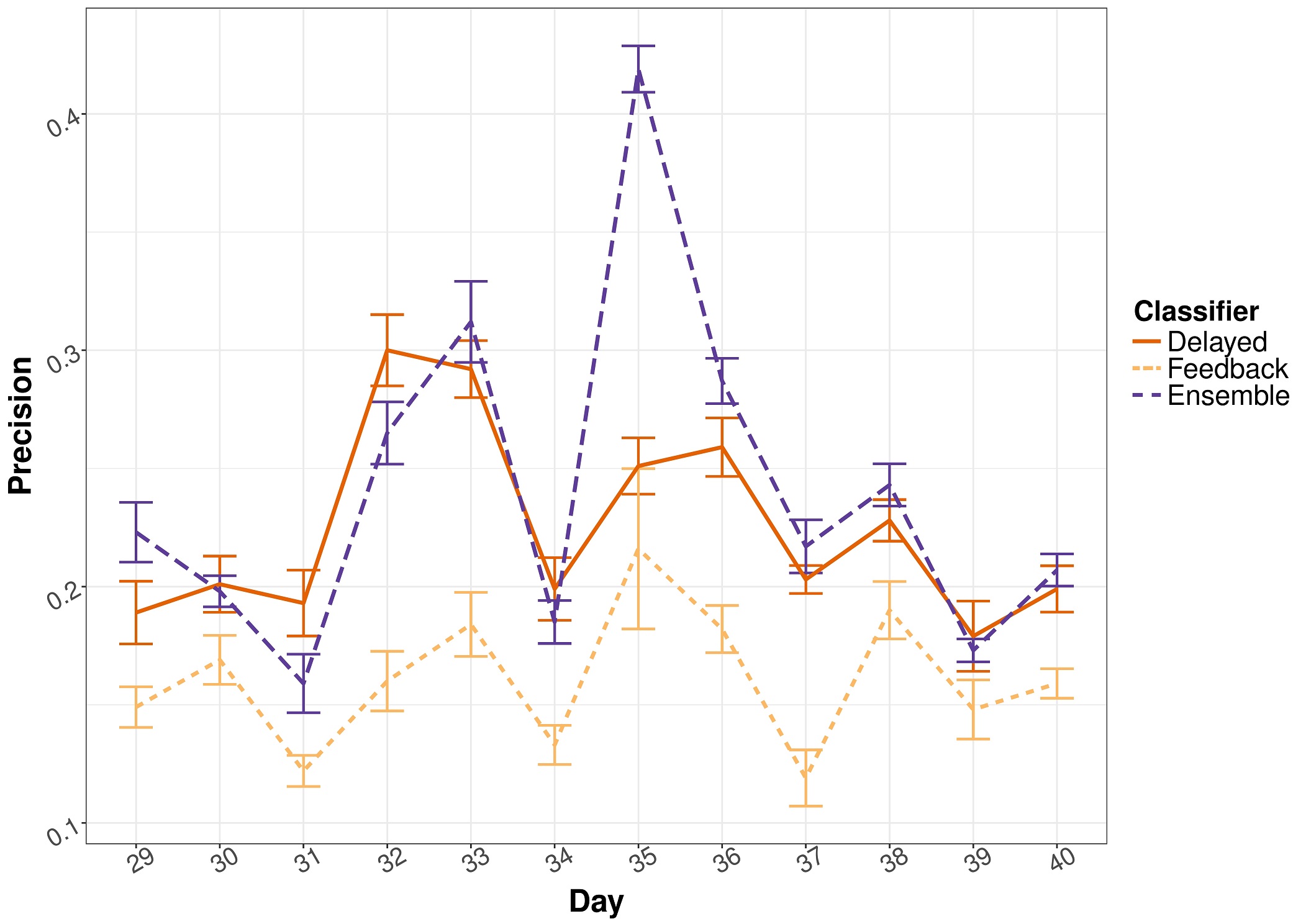}
    \caption{Percentage of detected frauds from the Delayed model, the Feedback model and the Ensemble of the two. Note that the aggregate model precision (dashed/violet line) is usually higher. The confidence intervals were obtained by repeating the experiment ten times and changing different random seeds.}
    \label{fig:accmod}
\end{center}
\end{figure*}

Fig.~\ref{fig:fraudsperday} reports the total number of fraudulent cards (solid/yellow line) and the percentage of detected fraudulent cards (dashed/violet line) for the analysed period.
As expected, we see an improvement of the precision when we move from the \emph{Partial ensemble} to the \emph{Full ensemble} phase, essentially due to the growing size of the classifier ensemble and the improvement of the Feedback model (initially trained on inaccurate alerts).
Fig.~\ref{fig:accmod} and Table \ref{table:acccpk} illustrate well the effectiveness  of the averaging strategy: most of the time and on average the single classifiers perform worse than the ensemble.
Those results have been obtained over ten runs of SCARFF, streaming the same time series, but changing the randomization seed.
We have also computed two paired t-tests between the $CP_k$ obtained using the Ensemble Classifier and those obtained by its two components.
The $CP_k$ of the Ensemble Classifier results to be statistically bigger than the Delayed and Feedback Classifier (p-values smaller than 10e-3).

\begin{table}[!ht]
\centering
\begin{tabular}{|l||r|}
 \hline
 \multicolumn{2}{|c|}{Fully Operational Stage} \\
 \hline
 Classifier & $CP_k$ \\
 \hline
 Delayed Classifier & 16.1\%\\
 Feedback Classifier  & 22.4\%\\
 Ensemble Classifier  & 24.1\%\\
 \hline
\end{tabular}
\caption{Precision $CP_k$ for multiple classifiers during the fully operational stage.}
\label{table:acccpk}
\end{table}

\begin{figure*}[!ht]
\begin{center}
    \includegraphics[width=1\textwidth]{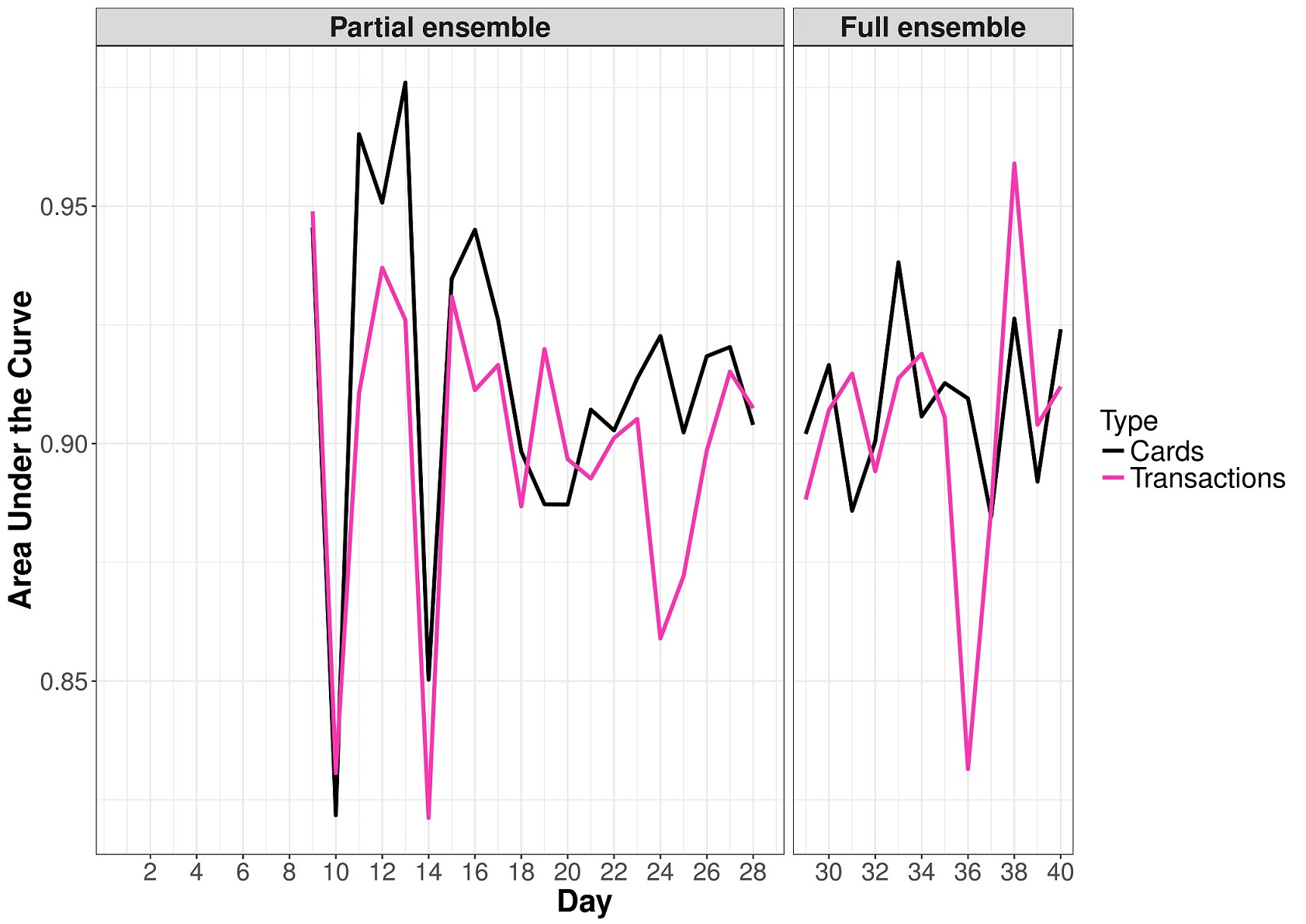}
    \caption{AUC of fraud detection at the level of card (darker color) and transaction (lighter color).}
    \label{fig:auc}
\end{center}
\end{figure*}

Since the $CP_k$ assessment refers to $k=100$ alerts only, we also report in Fig.~\ref{fig:auc} the \ac{AUC},which is a more general measure of accuracy used in fraud detection scenarios \cite{dal2014learned,van2015apate,Viaene04acase}. The two lines represent respectively the \ac{AUC} considering all the transactions and the most likely fraudulent transactions for each credit card.

Finally, a last measure of accuracy derived from the experiment relates to the capacity of our implemented model to detect a fraudulent card before what actually happened in the recorded dataset.
It happens in fact that some fraudulent cards were stopped only after several fraudulent transactions.
On the basis of our simulation, it appears that the implemented classifier is able to detect earlier fraudulent cards 3.6\% of the time.

\section{Conclusions and future work}\label{sec:C}

The paper presented SCARFF, an original scalable platform to automatically detect frauds in a near real-time horizon.
The most original contribution of this framework is the design and the implementation of an open source big data solution for real-world Fraud Detection and its test on a massive real-world data set. We wish to emphasize that the workflow proposed in our article, while not disclosing the data, has been made fully open source and reproducible by means of a \textbf{Docker}\footnote{\url{https://hub.docker.com/r/fabriziocarcillo/scarff/}} container and an artificial dataset. To the best of our knowledge this is the most complete and detailed open source and big data solution for credit card fraud detection in the literature. 



In terms of software development the paper shows that Kafka, Spark and Cassandra, may provide easy scalability and fault tolerance, to receive, aggregate and classify transactions at high rate.
In the experimental session we have extensively tested the system
in terms of scalability, sensitivity to parameters and
classification accuracy.

We have shown that the system behaves robustly up to an incoming rate of 200 transactions per second, which is a remarkable result once compared with the 2.4 transactions per seconds rate currently managed by our industrial partner.
Moreover, a rate of 200 transactions per seconds should not be considered as a hard upper limit since an appropriate setting of the number of executors and the batch duration could allow still higher rates.

In terms of precision we have confirmed previous results obtained in a conventional architecture with data resident in main memory.

Nevertheless,  as a concluding remark, it is important to add some words of caution about the maturity of big data solutions for large scale deployments.
Big data solutions are supported by a growing open-source community which leads to a very fast evolution and, at the same time, to a  high rate of new releases. If on the one hand this ensures rapid debugging, on the other hand it may induce instability in the existing running solutions. 
The problem is evident when one is trying to combine several functionalities of different tools in the same platform.
For instance we encountered several problems in querying a Cassandra table from Spark: we had a very hard time in doing ad hoc queries to Cassandra which had as consequence that we often decided to get the whole table from Cassandra and filter it on Spark (a suboptimal solution).

Overall, we consider that the most important message is that the adoption of a big data solution introduces a number of parameters having an impact on the resulting computational and classification performances.
In order to obtain an efficient solution to a specific detection problem, several trade-offs have to be made explicit and managed both at the software and hardware levels.
In our experience the most important trade-off concerned the number of transactions processed per second, the complexity of the feature engineering step, the batch duration and the number of available executors. 

Future work will focus on porting the existing solution to the industrial partner,  testing the efficiency in a Cloud environment, assessing the robustness to the adoption of alternative service providers (e.g. other databases than Cassandra) and generalizing the framework to other streaming settings (e.g. analytics of multivariate sensor streams).
From a more theoretical point of view, we would like to investigate innovative approaches especially in the area of semi-supervised and active learning.

\section*{Acknowledgement}
The authors FC, YLB and GB acknowledge the funding of the  Brufence project (Scalable machine learning for automating defense system) supported by INNOVIRIS (Brussels Institute for the encouragement of scientific research and innovation). ADP acknowledges the funding of the Doctiris (Adaptive real-time machine learning for credit card fraud detection) project supported by INNOVIRIS (Brussels Institute for the encouragement of scientific research and innovation).



\section*{References}
\bibliographystyle{elsarticle-num} 
\bibliography{biblio}





\end{document}